\documentclass[sigconf,nonacm]{acmart}
\renewcommand\footnotetextcopyrightpermission[1]{} 
\usepackage{todonotes}
\usepackage{changepage}
\usepackage{flushend}
\usepackage{fancybox,mdframed}
\usepackage{blindtext}
\usepackage{tikz}
\usetikzlibrary{arrows, shapes}
\usetikzlibrary{arrows.meta}
\usetikzlibrary{fit}
\usepackage{booktabs}
\usepackage{multirow}
\usepackage{subcaption}
\usepackage{booktabs}
\usepackage{colortbl}
\usepackage{graphicx}
\usepackage{balance}
\usepackage{float}

\begin{document}

\title{Recognizing Developers' Emotions while Programming}

\author{Daniela Girardi }
\orcid{0000-0002-4630-4793}
\affiliation{University of Bari, Italy}
\email{daniela.girardi@uniba.it}

\author{Nicole Novielli }
\orcid{0000-0003-1160-2608}
\affiliation{University of Bari, Italy}
\email{nicole.novielli@uniba.it}

\author{Davide Fucci }
\orcid{0000-0002-0679-4361}
\affiliation{Blekinge Institute of Technology, Sweden\\
University of Hamburg, Germany}
\email{davide.fucci@bth.se}

\author{Filippo Lanubile }
\orcid{0000-0003-3373-7589}
\affiliation{University of Bari, Italy}
\email{filippo.lanubile@uniba.it}

\begin{abstract}
Developers experience a wide range of emotions during programming tasks, which may have an impact on job performance. In this paper, we present an empirical study aimed at (i) investigating the link between emotion and progress, (ii) understanding the triggers for developers' emotions and the strategies to deal with negative ones, (iii) identifying the minimal set of non-invasive biometric sensors for emotion recognition during programming tasks. Results confirm previous findings about the relation between emotions and perceived productivity. Furthermore, we show that developers' emotions can be reliably recognized using only a wristband capturing the electrodermal activity and heart-related metrics.
\end{abstract}


\maketitle
\thispagestyle{empty}
\section{Introduction}
\label{sec:introduction}
Software development is an intellectual activity requiring creativity and problem-solving skills, which are influenced by affective states~\cite{GWA14}.
Previous work shows that positive emotions are beneficial for developers' well-being and productivity while negative ones lead to poor job performance and are detrimental to the software development process~\cite{GraziotinFWA18:JSS, Khan2011, MF15}.
Hence, emotion awareness---i.e, the awareness of own and others' emotions---is regarded as a key to success for software projects~\cite{Denning:2012}.

Early recognition of negative emotions, such as stress~\cite{MantylaADGO16}, frustration~\cite{FordP15:CHASE}, and anger~\cite{GachechiladzeLN17} can enable just-in-time corrective actions for developers and team managers, preventing burnout and undesired turnover~\cite{MantylaADGO16}.
Recent research findings demonstrate that negative emotions can be caused by uneven task distribution, wrong estimation of size and time required to complete an assignment,  difficulties solving a complex cognitive task, and obstacles when familiarizing with a new technology or programming language~\cite{FordP15:CHASE}.
Information about developers' emotional state can be leveraged to improve collaborative software development strategies~\cite{Guzman:2013}.
For instance, enriching retrospective meetings with feedback about developers' emotions can be used to reflect as a team on opportunities for improvement.
Thus, we envision the emergence and adoption of tools for enhancing emotion awareness during software development.

In this study, we focus on the identification of the emotions experienced by developers engaged in a programming task. Specifically, we operationalize emotions along the valence and arousal dimensions of the Circumplex Model of affect~\cite{JS1980}.
First, we build upon recent research investigating the relation between developers' emotions and perceived productivity~\cite{GraziotinWA15:Journal, MF15}.
We formulate our first research question as follows:
\begin{adjustwidth}{0.7cm}{}
\textit{RQ1} - What is the range of developers' emotions during a programming task and to what extent they correlate with their perceived progress?
\end{adjustwidth}
To address RQ1, we perform a study with 23 participants engaged in a programming task.
We ask participants to periodically self-report their emotional state and self-assessed progress.
We analyze the range of emotions reported and their correlation with perceived progress by fitting a linear-mixed effect model as done in the previous studies~\cite{MF15, GraziotinWA15:Journal}.

As a second goal, we aim at discovering the causes of positive and negative emotions experienced during software development.
Furthermore, we aim at identifying the coping strategies that might help programmers dealing with negative emotions.
As such, we formulate our second research question:
\begin{adjustwidth}{0.7cm}{}
\textit{RQ2} - What are the triggers for developers' emotions and the strategies they implement to deal with negative ones?
\end{adjustwidth}
To address RQ2, we interview the participants at the end of the study and perform manual coding~\cite{Martin:Turner} of their answers to open-ended questions.

Previous work demonstrated the feasibility of sensor-based emotion detection using non-invasive biometric devices~\cite{MF15, Girardi:ACII2017}.
However, the experimental setting usually employed in a laboratory setting is too complex for being applied in practice.
We aim at identifying the minimal set of sensors to wear in the work environment for reliable emotion recognition.
Accordingly, we formulate our third research question:
\begin{adjustwidth}{0.7cm}{}
\textit{RQ3} - What is the minimal set of non-invasive biometric sensors to recognize developers' emotions?
\end{adjustwidth}
We use supervised machine learning to train a classifier for developers' emotions based on biometric features, with different sensor configurations.

The contributions of this work are:
\begin{itemize}
    \item A list of emotional triggers related to software development, including strategies to deal with negative emotions.
    \item A set of supervised classifiers of developers' emotions. These include, to the best of our knowledge, the first attempt at classifying arousal during a programming tasks.
    \item A lab package\footnote{\url{https://figshare.com/articles/conference_contribution/Recognizing_Developers_Emotions_while_Programming/9206474}} to verify the results, replicate, and build upon this study.
\end{itemize}

The reminder of the paper is structured as follows.
Section~\ref{sec:background} describes the theoretical model we use to operationalize emotions, reviews  the existing literature on sensor-based emotion detection, and summarizes the empirical studies we build-upon in this work.
In Section~\ref{sec:design}, we describe the design of this study.
In Section~\ref{sec:analysis}, we report the analysis results and answer the research questions.
In Section~\ref{sec:discussion}, we compare our findings with previous work and discuss their implications and limitations.
Section~\ref{sec:conclusion} concludes the paper.

\section{Background}
In this section, we introduce the most important theories on modeling emotions and summarize the state of the art on recognizing emotions using biometric sensors. 
Moreover, we report details about two studies relating emotions to perceived progress in software engineering.  

\label{sec:background}
\subsection{Emotion Modeling}
\label{sec:modeling}
\begin{figure}[htb]
 \centering
\includegraphics[width=0.9\linewidth]{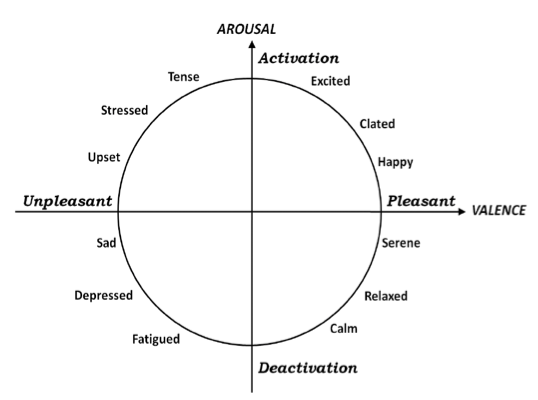}
  \caption{The Circumplex Model of Emotions~\cite{JS1980}.}
\vspace{-5mm}
\label{fig:russel}
\end{figure}

Psychologists worked on decoding emotions for decades, developing theories based on cognitive psychology and natural language communication.
Two theories have emerged. 
The first poses that a limited set of basic emotions exists. 
However, there is no consensus about their number or nature~\cite{Ekman1999,SLazarus1991-SLAEAA}. 
The second theory considers emotions as a continuous function of one or more dimensions~\cite{JS1980}. 
Dimensional models are not influenced by cultural or linguistic factors~\cite{guerini,Russel1991}, which makes them more robust compared to discrete models.
Consistently with prior research on emotion awareness in software engineering~\cite{MF15,GraziotinWA15:Journal,MantylaADGO16, Islam:2018:DSE:3167132.3167296}, we use a continuous representation of developers' emotions.  
Specifically, we refer to the Circumplex Model of Affect (see Figure~\ref{fig:russel}), which represents emotions according to two dimensions---\textit{valence} (pleasant vs. unpleasant) and \textit{arousal} (activation vs. deactivation).
According to this model, each emotion can be considered a ``label for a fuzzy set, defined as a class without sharp boundaries''~\cite{JS1980}. 
Pleasant emotional states, such as happiness, are associated with \textit{positive} valence, while unpleasant ones, such as sadness, are associated with \textit{negative} valence.
Arousal describes the level of activation of the emotional state ranging from inactive or \textit{low}, as in calmness or depression, to active or \textit{high}, as in excitement or tension.

\subsection{Sensor-based Emotion Classification}
\label{sec:sensorbased}
The link between emotions and physiological feedback---measured using biometric sensors---is investigated in the field of affective computing~\cite{KimA08,KoelstraMSLYEPNP12,SoleymaniAFP16}.
Among the several physiological measures that correlate with emotions, previous research investigated the electrical activity of the brain (EEG)~\cite{Kramer90physiologicalmetrics,Reuderink:2013,SoleymaniAFP16,Li:Lu:EEG}, the electrical activity of the skin (EDA)~\cite{Burleson:Picard, Kapoor:2007}, the electrical activity of contracting muscles (EMG)~\cite{KoelstraMSLYEPNP12,nogueira2013hybrid, Girardi:ACII2017}, and the blood volume pulse (BVP) from which heart rate (HR) and its variability (HRV) can be derived~\cite{Canento,Scheirer}. 

Specifically, changes in the EEG spectrum provide an indication of overall levels of arousal or alertness~\cite{Kramer90physiologicalmetrics} as well as pleasantness of the emotion stimulus~\cite{Reuderink:2013}. 
For example, Soleymani et al.~\cite{SoleymaniAFP16} found that high-frequencies sensed from electrodes positioned on the frontal, parietal, and occipital lobes have high correlation with valence.
Similarly, Li and Lu~\cite{Li:Lu:EEG} demonstrate that it is possible to discriminate between happiness and sadness based on the analysis of EEG signal. 

Concerning EDA, studies in psychology demonstrate how this signal considerably varies with changes in emotional intensity and specifically with the arousal dimension~\cite{BL2008}. 
Changes in EDA are a result of increased activity of the sweat glands, which takes place in presence of emotional arousal and cognitive workload. 
Hence, EDA has been employed to detect excitement, stress, interest, attention as well as anxiety and frustration~\cite{Burleson:Picard, Kapoor:2007}. 

BVP, HR, and HRV metrics---captured by a plethysmograph---have been successfully employed for emotion recognition~\cite{Canento,Scheirer}. 

Facial EMG is particularly useful in predicting emotions~\cite{KoelstraMSLYEPNP12,nogueira2013hybrid}. However, its usage leads to poor results when the sensors are placed on body parts other than the face, such as the arms~\cite{Girardi:ACII2017}.
Accordingly, we exclude EMG from this study.  

In this study, we include measures from EEG, EDA, BVP, and HR as they can be collected using low-cost noninvasive sensors~\cite{Girardi:ACII2017,MF15} that can be comfortably used by developers during programming tasks (see Section~\ref{sec:instrumentation}). 
This choice is in line with current research investigating the use of lightweight biometric sensor for studying human aspects in software development. 
Fucci et al.~\cite{FucciGNQL19} use EEG, EDA, and heart-related measurements for the automatic identification of code comprehension tasks.
Fritz et al.~\cite{FritzBMYZ14} rely on a combination of EEG, BVP, and eye tracker to assess difficulty in code comprehension and prevent developers from introducing bugs.
In a follow-up study, they employ the same set of sensors to distinguish between positive and negative emotions during programming tasks~\cite{MF15}. 
Similarly, EDA, HR, HRV, and breath-related metrics have been used in a field study to identify code quality concerns during software development~\cite{MullerF16}.
Z\"ueger et al.~\cite{ZugerMMF18} combine heart-related metrics with a wristband activity tracker to predict developers' interruptibility.

\subsection{Former Studies}
\label{sec:relatedstudies}
Our study builds upon the design and results of two former studies linking developers' emotions with their perceived progress ~\cite{GraziotinWA15:Journal, MF15}.
They follow similar experimental protocols involving a longitudinal study with repeated measures of novice and professional developers' emotional states.
In particular, Graziotin et al.~\cite{GraziotinWA15:Journal} recruited eight subjects of which four professional developers and four undergraduate students with major in computer science, whereas M\"uller and Fritz~\cite{MF15} observed 17 subjects of which six professionals with an average experience of seven years and 11 PhD students in computer science. 

In both studies, the participants were observed during a programming session and interrupted every five minutes to answer a self-report survey. 
Emotions were self-reported along the valence and arousal dimensions. 
M\"uller and Fritz measured the self-perceived participants' progress while completing two programming tasks (30 minutes each).
Conversely, Graziotin et al. asked participants to report on perceived productivity while working for 90 minutes in a natural setting---i.e., on their own projects.
Graziotin et al. also measured dominance---i.e., the extent to which a subject feels in control or controlled~\cite{BradleyLang94}.



The studies of Graziotin et al.~\cite{GraziotinWA15:Journal} and M\"uller and Fritz~\cite{MF15} provide empirical evidence that valence correlates with self-perceived productivity and progress, respectively.
Among the main causes for negative emotions, leading to the perception of being stuck, M\"uller and Fritz report cognitive difficulties, impossibility to fulfill information needs, and code not working. 
Conversely, being able to understand the code and identify a solution strategy are among the top reasons for positive affect.

Finally, M\"uller and Fritz trained a supervised emotion classifier able to distinguish between positive and negative emotions with an accuracy of $71\%$. 
In their setting, they use multiple sensors including EEG, EDA, HR, and eye tracking metrics. However, they neglect the arousal classification.


\section{Study Design}
\label{sec:design}
In this section, we report a brief characterization of the participants in our study, the study task, the tools and devices used to measure the relevant constructs, and the study protocol.

\subsection{Participants}
We recruited 27 CS students (23 males, four females) from the Department of Computer Science of our University of which 21 undergraduates, five graduates, and one post-graduate. 
Following a convenience sampling strategy, we recruited volunteers as participants only if they could provide evidence they cleared exams where Java programming (e.g., the programming language for the this study task) was used for capstone projects. 

\subsection{Development Task} We use one of the two tasks, including the materials, designed by M\"uller and Fritz in their study~\cite{MF15}.
The task consisted in writing a Java program using the StackExchange API \footnote{https://api.stackexchange.com} to retrieve all answers posted by a specific user on StackOverflow and sum up the scores the user earned for these answers.
The participants were provided with a skeleton code which they had to modify to complete the task. 

\subsection{Measurement Tools and Devices}
\label{sec:instrumentation}
\begin{figure}[htb]
    \centering
    \includegraphics[width=0.8\linewidth]{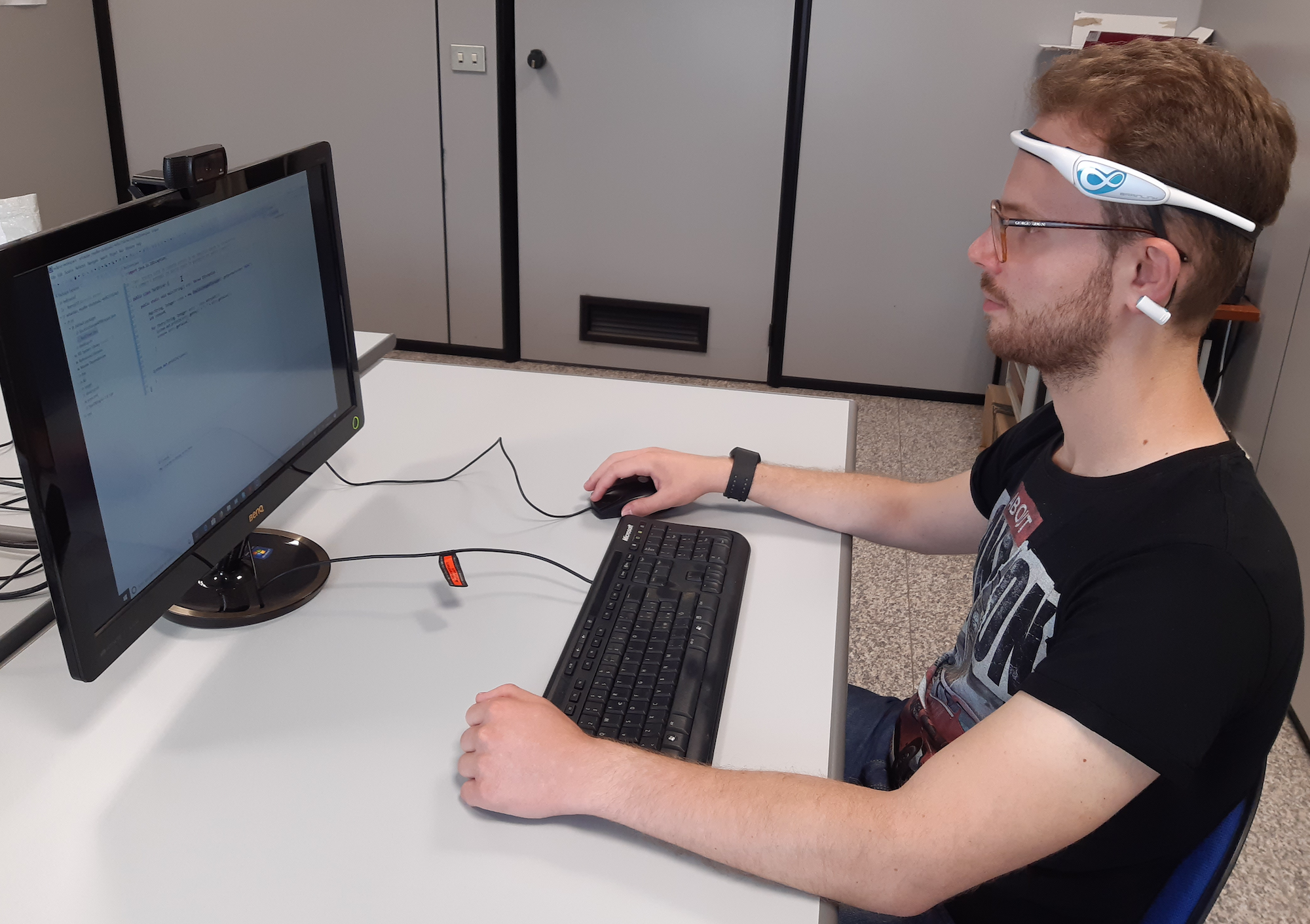}
    \caption{A participant wearing the Empatica E4 wristband and the BrainLink headset during the study.}
    \label{fig:exp_session}
\end{figure}

\textit{\textbf{Biometric sensors.}} 
We measure the subjects' physiological signals using lightweight biometric sensors analogous to those employed by M\"ueller and Fritz~\cite{MF15}---i.e., comfortable to wear in the work environment \cite{Girardi:2019}.
Specifically, we use the NeuroSky BrainLink headset to record the EEG waves and the Empatica E4 wristband for EDA, BVP, and heart-related metrics (see Fig.~\ref{fig:exp_session}).

The BrainLink EEG uses one electrode placed on the surface of the scalp and a reference placed on the earlobe\footnote{The reference electrode is used to compute the amount of information of the active EEG electrode placed on the forehead. As such, it has to be placed on a neutral place like the earlobe~\cite{EEG:reference}} (see Figure~\ref{fig:exp_session}).
EEG waves are extracted by pre-processing the raw signal collected by the device with a sample frequency of 512Hz (see Section~\ref{sec:rq3}).
Besides raw EEG signal, BrainLink extracts metrics related to mental focus (i.e., \textit{attention}) and calmness (i.e, \textit{meditation}).~\footnote{\url{http://developer.neurosky.com/docs/doku.php?id=esenses_tm}}

The Empatica E4 wristband measures EDA with a sample frequency of 4Hz.
It features a plethysmograph for collecting BVP sampled at a frequency of 64Hz.
BVP is used to derive the HR and HRV. 
Following the guidelines provided by the Empatica, support\footnote{https://support.empatica.com/hc/en-us/articles/360030058011-E4-data-IBI-expected-signal} we decided to exclude HRV as it is not reliable in dynamic conditions (i.e., while typing). 

\begin{figure}
    \centering
    \includegraphics[width=\linewidth]{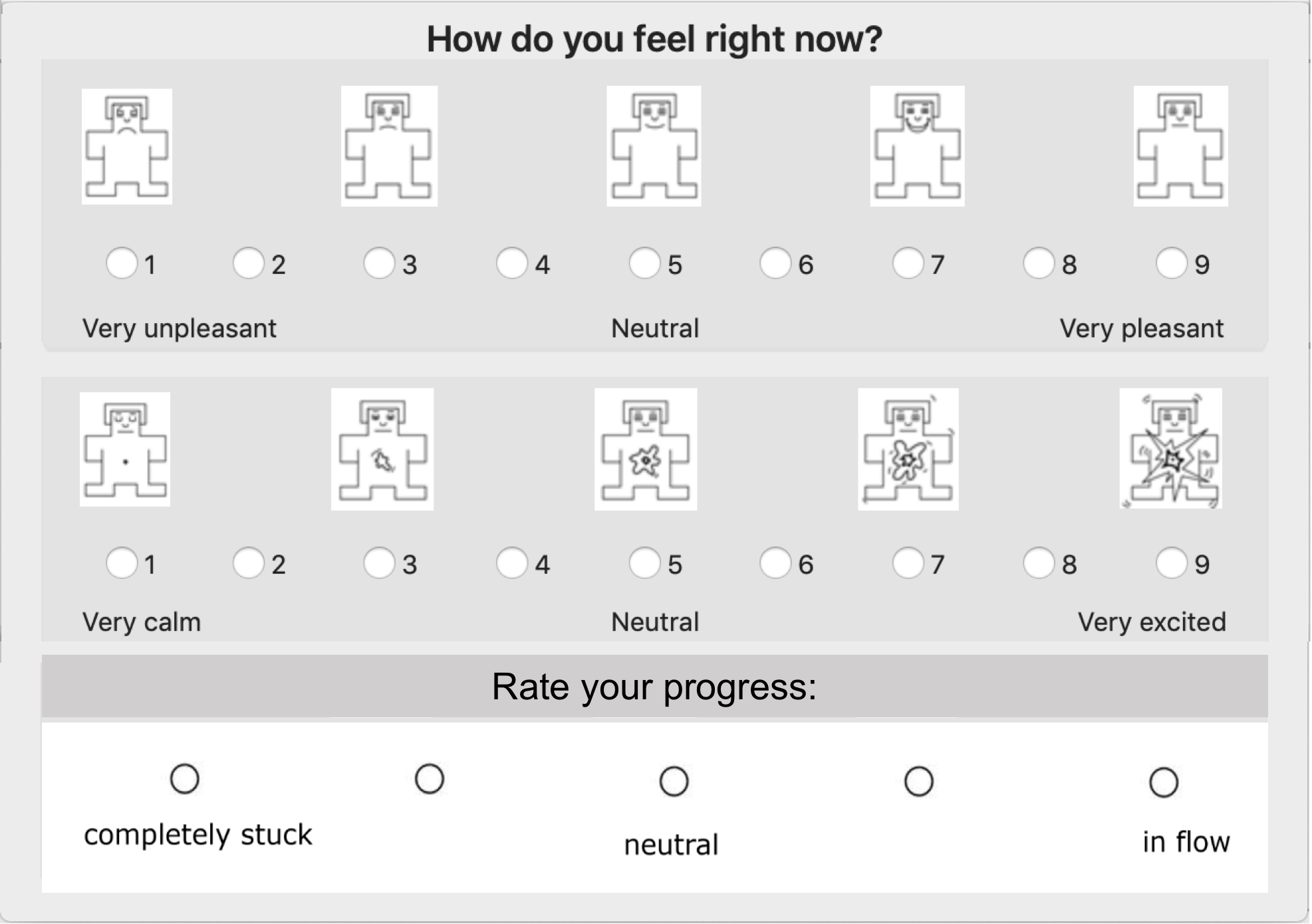}
    \caption{The SAM mannequins for assessment of valence and arousal and the 5-point Likert scale question for assessment of the perceived progress.}
    \label{fig:SAM}
\end{figure}

\textit{\textbf{Self-report of Emotions and Progress.}} The measurement of emotions and perceived progress is performed through experience sampling~\cite{GraziotinWA15:Journal, MF15}.
In line with the approach implemented by Graziotin et al.~\cite{GraziotinWA15:Journal}, we collect self-reported valence and arousal ratings during interruptions of the development task using Self-Assessment Manikin (SAM)~\cite{BradleyLang94}. 
Consistently with previous assessment of valence and arousal in affective computing research~\cite{KoelstraMSLYEPNP12}, we use a nine-point scale.
Figure~\ref{fig:SAM} shows the SAM mannequins for valence and arousal (top) as well as a 5-point Likert item (bottom) to assess the perceived progress.  
During the interruptions, we prompt the participant to elaborate on the triggers for the emotional state and take notes of their answers. 

\textit{\textbf{Debriefing questionnaire.}} 
To elicit triggers and strategies for handling emotions, the first author interviews each participant at the end of the development task, asking the following questions:
\begin{itemize}
    \item What are the causes for positive emotions during programming?
    \item What are the causes for negative emotions during programming? \item Which strategies do you adopt to deal with negative emotions?
\end{itemize}

\subsection{Experimental Protocol}
~\label{sec:protocol}
\begin{figure*}[htb]
    \centering
    \includegraphics[width=0.80\linewidth]{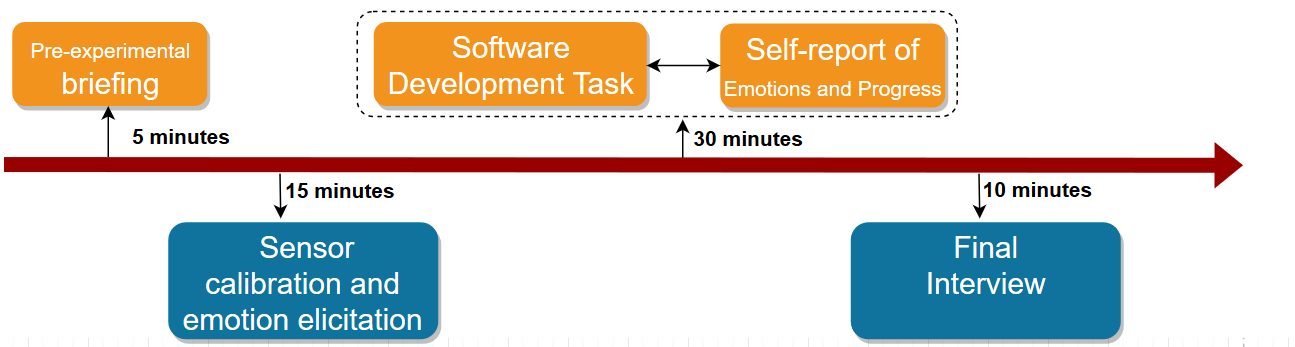}
    \caption{The timeline of the study.}
    \label{fig:timeline}
\end{figure*}

We organize the study according to the following phases, as shown in Figure~\ref{fig:timeline}.

\textit{\textbf{Pre-experimental briefing.}} The participant gets acquainted with the settings---e.g., sitting in a comfortable position, adjusting the monitor height. 
The experimenter summarizes the upcoming steps and explains the programming task. 
The participant signs the consent form to allow anonymous treatment of the collected data. 

\textit{\textbf{Sensor calibration and emotion elicitation.}} The participant wears the biometric sensors (see Figure~\ref{fig:exp_session}) and the experimenter checks  that the devices record the signals correctly.

Before the participant starts working on the actual programming assignment, the experimenter asks her to take part in an emotion-elicitation task.
The purpose of this step is two-fold.
On one hand, it allows the subject to get acquainted with the SAM mannequins; on the other hand, it allows the experimenter to collect her biometric and SAM feedback, both in a neutral condition and in presence of controlled  stimuli. 
This step follows the design described in a previous emotion-elicitation study~\cite{Girardi:ACII2017}.
The participant watches eight videos, selected from the DEAP dataset~\cite{KoelstraMSLYEPNP12}, associated with valence and arousal scores on a scale from 1 to 9.
Each video is mapped to the four quadrants of the emotional space in the Circumplex Model of Affect based on a discretization of the scores. 
Specifically, the eight videos used in the emotion elicitation are equally distributed among the four quadrants---i.e., $positive$ valence and $high$ arousal, $positive$ valence and $low$ arousal, $negative$ valence and $low$ arousal, $negative$ valence and $high$ arousal. 
The emotion elicitation step lasts 10 minutes, with the eight videos presented in four sessions.
Each session consists of a 30-second baseline video showing a quiet image with relaxing music in the background, followed by a 2-minute display of the selected videos (one minute per video). 
At the beginning of each trial, a 3-second screen displays the current trial number to make the participant aware of her progress. 
After each video, the participant is instructed to report her emotions using the SAM mannequins. 
Therefore, each subject provided sixteen pairs of  ratings---i.e., one valence and one arousal assessment per video. 

Prior to the development task, the participant watches a 2-minute relaxing video of a nature scenery to induce relaxation and a neutral emotional state~\cite{rottemberg:2007}.
We use biometrics recorded when showing such video as physiological baselines for the participants. 

\textit{\textbf{Software development task and self-report.}} The core of the study is a 30-minute development session during which we apply experience sampling~\cite{Larson2014}.
The experimenter (i.e., the first author) observed the behavior of participants during the entire session and interrupted them every five minutes, asking to report their emotions, perceived progress, and to provide information about the reasons for their emotions.
We choose a time frame of five minutes as it represents the average time for which developers stay focused on a single task~\cite{Meyer:2014,MF15}. 
We collect the subjects' biometrics during the entire duration of the development task.
After the task, participants watch again the 2-minutes relaxing video to ward-off possible induced emotions, for example, from not succeeding in solving the task.
In total, each participant provides six ratings for valence and six for arousal (i.e., one valence-arousal rating for each interruption).
In addition, participants provide six progress ratings (one per interruption). 
Finally, they provide six answers to the open-ended question about the triggers for reported emotions, which we use in the data quality assurance step.

\textit{\textbf{Final interview}}. We run the post-experimental debriefing with each participant for approximately 10 minutes.
The experimenter interviews participants to investigate i) the triggers for positive and negative emotions during the task, and ii) the strategies subjects implement to deal with negative emotions. 
The participants could also ask questions and give feedback about the experiment.
Finally, participants are rewarded with a voucher for a meal.

\textit{\textbf{Data Quality Assurance}}. Once the experiment was completed, but before analyzing the data, we manually performed a sanity check of the collected data. 
In particular, we investigated potential malfunctioning of the sensors which can introduce noise and discontinued recording of raw signals. 
In addition, we check the consistency of the self-reported valence, arousal, and progress scores with respect to the comments provided in the open questions.
We looked for signs of negligence, inconsistencies, and misinterpretation of the guidelines for using the SAM-based report. 
For example, one subject scored his arousal as eight for all the interruptions, however, this did not match the content of his comments. 
As a results of this step, we discarded four participants (all males). Our final pool of participants include 23 subjects overall, of which four females.


\section{Analysis and Results}
\label{sec:analysis}

In this Section, we answer our research questions by analyzing the data collected in the study using a mix of quantitative and qualitative methods.

\subsection{Experienced Range of Emotions and their Correlation with Progress (RQ1)}
\label{sec:rq1}
We analyze the self-reported scores for emotions and progress collected through experience sampling. As described in Section~\ref{sec:protocol}, each participant provided self-assessment of emotions at each interruption, thus reporting six pairs of SAM-ratings for valence and arousal. As such, our dataset includes $138$ ($6x23$ subjects) ratings for each emotion dimension. Analogously, we collected $138$ scores ($6x23$) for progress. 

To investigate the range of developers' emotions during the programming task, we following the approach proposed by M\"ueller and Fritz~\cite{MF15}. We analyze and compare the emotion scores the developers reported during the programming task and emotion elicitation.   

\begin{figure}[htb]
    \centering
    \includegraphics[width=\linewidth]{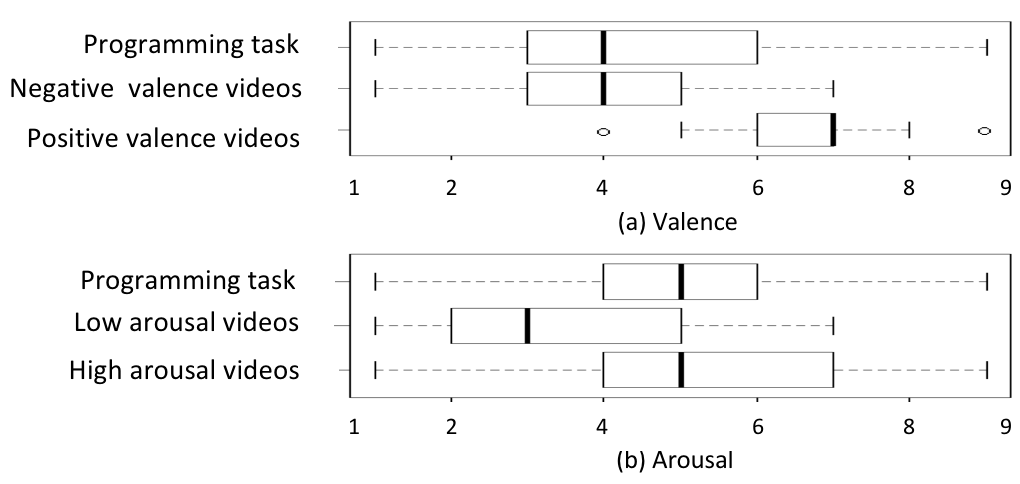}
    \caption{Box plots for valence (a) and arousal (b) during the software development task and emotion elicitation video.}
    \vspace{-3mm}
    \label{fig:boxplot}
\end{figure}

Figure~\ref{fig:boxplot} reports the SAM scores for valence and arousal.
We use the scores provided during the video-driven emotion elicitation as a reference for the way participants report valence and arousal in presence of the controlled stimulus.
For example, the range of SAM scores reported for positive videos indicate how the participants rate their valence when experiencing a positive emotion. 
We observe that the entire range of emotions is covered by the scores reported while watching videos as well as programming. 
For valence, we observe a clear distinction between the ranges used for positive and negative videos (Figure~\ref{fig:boxplot}a).
In addition, negative emotions tend to prevail during the programming task---i.e., the median score corresponds to the one reported for the negative videos. 
Conversely, the median value for the arousal distribution of the programming task (Figure~\ref{fig:boxplot}b) corresponds to the one reported for high arousal videos.

\begin{table}[]
\begin{tabular}{lll|l|ll}
 & \multicolumn{2}{c}{\textbf{Stuck}} & \multicolumn{1}{c}{\textbf{Neutral}} & \multicolumn{2}{c}{\textbf{In flow}} \\ \hline
 & \textit{Score 1} & \textit{Score 2} & \textit{Score 3} & \textit{Score 4} & \textit{Score 5} \\
\cline{2-6} 
& 28 (20\%) & 59 (43\%) & 36 (26\%) & 11 (8\%) & 4 (3\%) \\
\multicolumn{1}{r}{\textit{overall}} & \multicolumn{2}{c}{87 (63\%)} & \multicolumn{1}{c}{36 (26\%)} & \multicolumn{2}{c}{15 (11\%)}\\ \hline
\end{tabular}
\caption{Frequency of progress scores on a 1--5 scale (n=138).}
\label{tab:progressScores}
\end{table}

The participants report the whole range of progress (i.e., 1--5), from \textit{completely stuck} (score equal to 1) to \textit{in flow} (score equal to 5), with median = 3 and inter-quartile range = 2 (see Table~\ref{tab:progressScores}). 
Most developers had troubles solving the task---only four succeeded in developing a complete solution. 
Accordingly, most of the time they reported being stuck (63\% of answers).
They reported being in a neutral state 26\% of the cases, corresponding to 36 answers, and \textit{in flow} in 11\% of the cases (15 answers).

We investigate the link between reported emotions and perceived progress by fitting a linear mixed model, which is robust in case of repeated measurements and longitudinal data~\cite{Gueorguieva2004}.
To create the model, we used the \texttt{lme4} R package\footnote{https://cran.r-project.org/web/packages/lme4/index.html}.
Consistently with the approach adopted in the former studies~\cite{MF15,GraziotinWA15:Journal}, we consider progress as the dependent variable and valence, arousal, and their interaction with time as fixed effects.
Given our study design, we cannot exclude that the perceived progress can be impacted by time. 
Therefore, time and its interaction with the emotional dimensions are also included in the model.
To account for individual differences in the SAM reports, we standardize the valence and arousal using Z-scores~\cite{MF15, GraziotinWA15:Journal}. 

\begin{table}[htb]
\begin{tabular}{l|l|p{1.1cm}|p{1.1cm}|p{1.3cm}}
\textbf{Fixed Effects} & \textbf{Estimate} & \textbf{Upper p-value} (132 d.f.) & \textbf{Lower p-value }(103 d.f.) & \textbf{Dev. explained}\\
\hline
Valence       & 0.17 (*) &     0.00          &     0.00          & 27.8\% \\
Arousal       & -0.05     &      0.23         &      0.23         & 0.5\% \\
Time          & -0.02     &      0.97         &   0.97            & 0.0\% \\
Valence:Time  & 0.03 (*)     &      0.05         &    0.05           & 1.4\% \\
Arousal:Time  & 0.02     &       0.24        &     0.25          & 0.48\% \\      
\hline
\end{tabular}
\caption{Parameter estimation for the fixed effects on Perceived Progress ($*$ indicates a significant estimate with $\alpha = 0.05$).}
\vspace{-3mm}
\label{tab:mixedmodel}
\end{table}

In Table~\ref{tab:mixedmodel}, we report the parameter estimation for the mixed-effect model and the percentage deviance explained by each effect.
Our model significantly differs from the null model---i.e., the model with no correlation between fixed effects and progress ($\chi^2(5) = 63, p<0.001$).  
We observe a significant effect of valence on progress at 95\% confidence level.
Valence also shows the highest explanatory power with $27.80\%$ of deviance explained, compared to $30.15\%$ observed for the whole model. 
Conversely, we did not observe any effect of arousal on self-reported progress. 
This result holds for the effect of time and its interaction with arousal. 
Our model shows a statistically significant correlation between progress and the interaction between time and valence. 
However, the effect of such interaction on the overall model is small (1.4\% of deviance explained). 


\begin{mdframed}[backgroundcolor=lightgray!20]
\textit{Summary RQ1} - Developers experience a wide range of emotions during programming tasks. We observe a prevalence of negative valence and high arousal. Valence is positively correlated with perceived progress. 
\end{mdframed} 

\subsection{Triggers and Strategies for Emotions (RQ2)}
\label{sec:rq2}
To investigate triggers for emotions, as well as the strategies to deal with negative ones, we manually analyzed the 69 answers provided during the debriefing questionnaire in the final interview, -- i.e., three for each participant.

We performed qualitative data analysis using a sentence-by-sentence approach in a semi-exploratory fashion. 
We applied selective coding~\cite{Martin:Turner} based on the constructs associated with the research question (i.e., positive and negative emotions, as well as strategies for coping with the latter).
We identified 29 sentences discussing positive emotions triggers, 41 for negative emotions triggers, and 47 for coping strategies. 
Subsequently, two researchers coded each sentence following an open coding approach~\cite{Martin:Turner}. 
During a meeting, the researchers reconciled their codes in a single one.
We obtained 23 codes---eight reasons for positive emotions, eight for negative ones, and seven strategies for dealing with negative emotions.
These codes were then grouped to form relationships and themes captured by applying axial coding~\cite{Martin:Turner}.
Three themes emerge: \textit{self} refers to the developers' dimension, \textit{social} refers to peers and collaborators, and \textit{solution} refers to issues with artifacts, design, and implementation of the task. 

\begin{table*}[htb]
\begin{tabular}{lllll}
\hline
&  & \textit{\begin{tabular}[c]{@{}l@{}}\textbf{Self}\\ \textit{11 codes (43)}\end{tabular}} & \begin{tabular}[c]{@{}l@{}}\textbf{Social}\\ \textit{15 codes (19)}\end{tabular} & \begin{tabular}[c]{@{}l@{}}\textbf{Solution}\\ \textit{9 codes (25)}\end{tabular}\\
\hline
\multirow{2}{*}{Triggers} & \begin{tabular}[c]{@{}l@{}}\textbf{Positive Emotions}\\ \textit{8 codes (24)}\end{tabular} & \begin{tabular}[c]{@{}l@{}}New challenges (3)\\ In flow (3)\\ Personal experience (1)\end{tabular} & \begin{tabular}[c]{@{}l@{}}Knowledge sharing (2)\\ Collaborative development (1)\end{tabular} & \begin{tabular}[c]{@{}l@{}}Incremental (8)\\ Simplicity (4)\\ No errors (2)\end{tabular} \\
\cline{3-5}
 & \begin{tabular}[c]{@{}l@{}}\textbf{Negative Emotions}\\ \textit{8 codes (31)}\end{tabular} & \begin{tabular}[c]{@{}l@{}}Being stuck (8)\\ Time pressure (6)\\ Multitasking (1)\end{tabular} & \begin{tabular}[c]{@{}l@{}}Unavailable collaborators (3)\\ Peer pressure (2)\end{tabular} & \begin{tabular}[c]{@{}l@{}}Unexpected output (4)\\ Unexpected usage (4)\\ Unavailable documentation (3)\end{tabular} \\
 \hline
Strategies & \begin{tabular}[c]{@{}l@{}}\textbf{Negative Emotions}\\ \textit{7 codes (32)}\end{tabular} & \begin{tabular}[c]{@{}l@{}}Take breaks (17)\\ Change task (1)\\ Start over (1)\\ Decompose problem (1)\\ Change approach (1)\end{tabular} & \begin{tabular}[c]{@{}l@{}}Look for collaboration (9)\\ Explain problem (2)\end{tabular} & --- \\
\hline
\end{tabular}
\caption{Themes identified after coding. Code occurrences are reported in parentheses.}
\label{tab:coding}
\vspace{-3mm}
\end{table*}

Table~\ref{tab:coding} shows the themes identified as the result of the coding process. 
The most frequent trigger for emotions refers to the \textit{solution} dimension (25 occurrences overall, of which 14 for positive and 11 for negative). 
The participants felt particularly happy when able to \textit{incrementally} implement the designed solution and when  believing the solution itself is \textit{simple}.
Analogously, \textit{unexpected output}, \textit{unexpected usage} of libraries, and \textit{unavailable documentation} trigger negative emotions.
The most frequent causes for negative valence relate to the self dimension (15 occurrences), with the developers reporting the feeling of \textit{being stuck} and the awareness of \textit{time pressure} as the main causes for negative emotions. 
Finally, the \textit{social} theme appears with a low frequency as the participants had to complete the programming task by themselves. 

As for strategies developers implement to deal with negative emotions, \textit{take breaks} is the most popular one, immediately followed by \textit{look for collaboration}---e.g., asking help from peers.
Some participants also report \textit{changing task} and \textit{starting over} as strategies to regain focus and shift towards  positive emotions.
Similarly, they indicated \textit{changing approach} to the solution and \textit{decomposing the problem} into simpler ones as strategies to gain confidence and react to negative emotions. 

\begin{mdframed}[backgroundcolor=lightgray!20]
\textit{Summary RQ2} - Developers' positive emotion are mainly triggered by the  effectiveness of the implemented solution. Unexpected code behavior and missing documentation cause negative emotions, which are also due to time pressure and self-perceived low productivity. To deal with negative emotions, developers take breaks and look for peers' help.
\end{mdframed} 

\subsection{A Minimal Set of Sensors for  Classifying Developers' Emotions (RQ3)}
\label{sec:rq3}
\begin{figure}[htb]
    \centering
    \includegraphics[width=\linewidth]{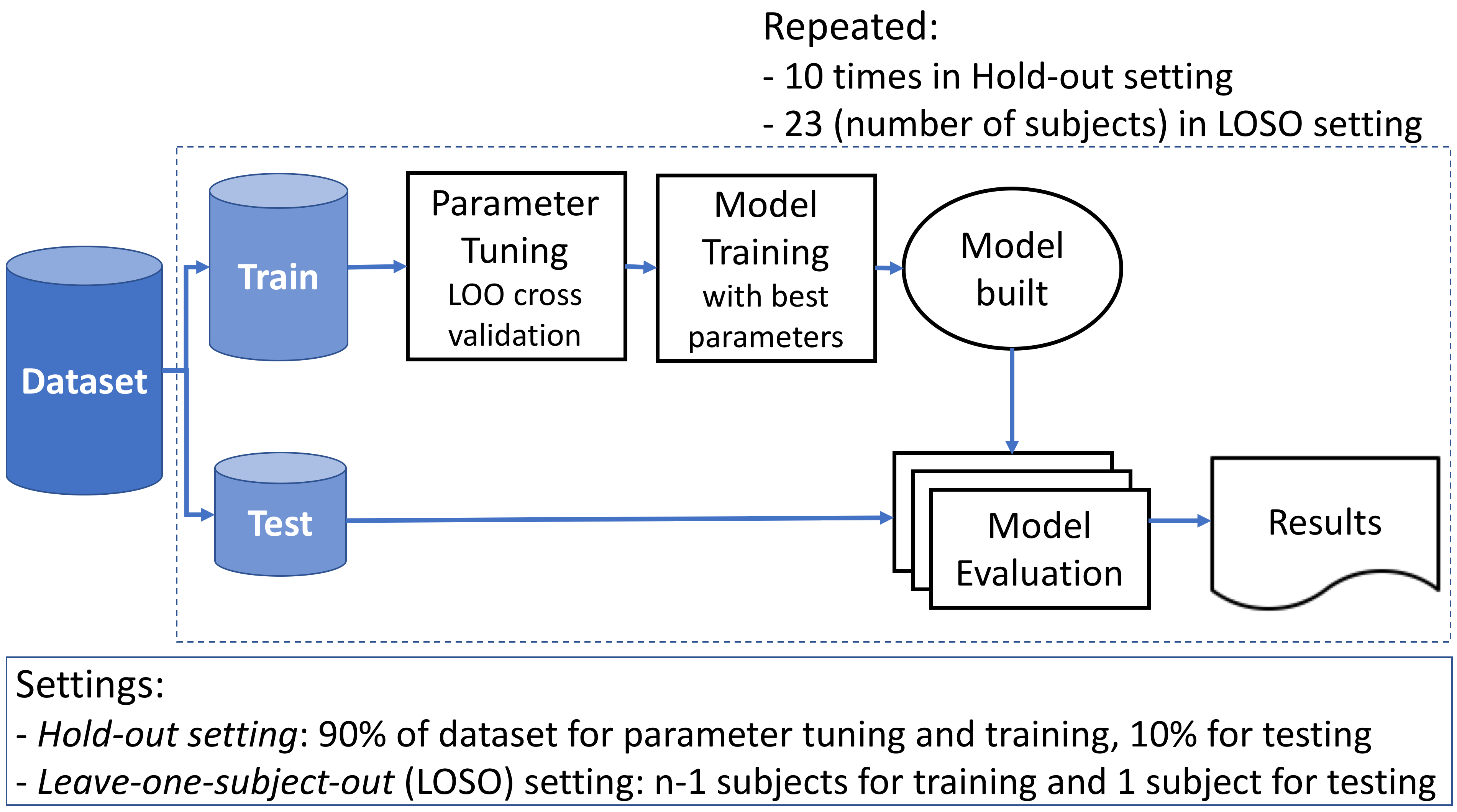}
\caption{The machine learning pipeline implemented, with the Hold-out and Leave-one-subject-out evaluation settings.}
    \label{fig:expSetting}
\end{figure}

We address RQ3 using a machine learning approach for classifying the participants' emotions during the development task.
Fig.\ref{fig:expSetting} shows the machine learning pipeline we implemented to answer this research question. 

\label{sec:machine_learning}
\textit{\textbf{Dataset}} The dataset consists of the self-reported emotions of participants during the development task.
Each of the 23 participants performed a total of six SAM-based assessments of valence and arousal. 
As a result, we obtained two datasets of 138 observations, one for valence and one for arousal.
Accordingly, we trained two separate classifiers by considering features extracted from the biometric signals associated to each observation, as captured by the sensors. 

We define the \textit{positive} and \textit{negative} labels for valence, and \textit{high} and \textit{low} labels for arousal.
For this purpose, we discretize the SAM scores following the approach applied by M\"uller and Fritz~\cite{MF15}.
First, we adjusted the valence and arousal scores based on the mean values reported while watching the emotion-triggering videos. 
Following such approach, we defined gold labels for valence and arousal by taking into account (and correcting for) fluctuations due to the participants' subjective interpretation of the SAM scale.
Then, we assigned a \textit{positive} valence label (respectively a \textit{high} arousal label) to instances with scores above the mean and a \textit{negative} valence label (respectively a \textit{low} arousal label) to instances with scores below it. 
Finally, following the recommendation reported in previous work~\cite{MF15}, we manually inspected 31 observations for valence and 30 for arousal for which we observed scores in the $mean \pm0.5$ interval. 
For such cases, two authors manually assigned valence and arousal labels.
They obtained a substantial agreement~\cite{Viera:Garret}, with $\kappa = .67$ and observed agreement $= 84\%$. 
The few disagreement cases (less than five for each dimension) were resolved in a discussion, following a consolidated approach in affective computing research~\cite{Basile2018}.
At the end of this process, we obtained the distribution reported in Table~\ref{tab:goldstandard}.

\begin{table}[]
\begin{tabular}{lcc|lcc}
\hline
\multirow{2}{*}{\textit{Valence}} & \textbf{Positive} & \textbf{Negative} & \multirow{2}{*}{\textit{Arousal}} & \textbf{High} & \textbf{Low} \\
 & 44 (32\%) & 94 (68\%) &  & 85 (62\%) & 53 (38\%)\\
 \hline
\end{tabular}
\caption{Gold standard for valence and arousal.}
\vspace{-7mm}
\label{tab:goldstandard}
\end{table}

\textit{\textbf{Preprocessing and Features extraction}} Although the biometric signals were recorded during the entire experimental session for all the participants, we only consider the signals recorded in proximity of the stimuli of interest---i.e., the signals collected in the 10 seconds before the subjects were interrupted.
This choice is in line with consolidated practices in related research on sensor-based classification of emotional~\cite{MF15,Girardi:ACII2017} and cognitive states~\cite{FucciGNQL19} of software developers.
To synchronize the measurement of the biometric signals with the emotion self-assessment, we \textit{(i)} save the timestamp of the interruption \texttt{(t\_interruption)}, 
 \textit{(ii)} calculate the timestamp for relevant timeframe for each interruption---i.e., 10 seconds before the self-report \texttt{(t\_start)}, and 
   \textit{(iii)} select each signal samples recorded between \texttt{t\_start} and \texttt{t\_interruption}.


For each participant, we normalize the signals to her baseline using Z$-$score normalization \cite{MF15}. 
The baseline is calculated considering the last 30 seconds of the video used to elicit a neutral state before starting the task~\cite{FritzBMYZ14}. 

To maximize the signal information and reduce noise caused by movements, we applied multiple filtering techniques.
Regarding EEG and BVP, we extract frequency bands using a band-pass filter algorithm  at different intervals~\cite{Canento}.
The EEG signal can be decomposed into five waves based on the frequency, namely delta ($<$ 4Hz), theta (4-7,5Hz), alpha (4-12,5Hz), beta (13-30Hz), and gamma ($>$ 30Hz).
We apply the filter to extract the distinct cerebral waves as each spectrum might provide different information.
The EDA signal is constituted by a tonic component, indicating the level of electrical conductivity of the skin, and a phasic component, representing the phasic changes in electrical conductivity or skin conductance response (SCR)~\cite{BWJR15}. 
We applied the cvxEDA algorithm~\cite{GVLSC16} to extract the two components. 

\begin{table}[htb]
\begin{tabular}{p{0.6cm}p{7cm}}
\textbf{Signal} & \textbf{Features} \\
\hline
\multicolumn{2}{c}{\textit{Sensor: Brainlink}} \\
\hline
EEG & \begin{tabular}[c]{@{}l@{}}- Frequency bin for alpha, beta, gamma, delta, theta waves\\ - Ratio between frequency bin of each band \& one another\\ - For attention and meditation: min, max, \\ difference between mean for baseline and task\end{tabular} \\
\hline
\multicolumn{2}{c}{\textit{Sensor: Empatica E4}} \\
\hline
EDA & \begin{tabular}[c]{@{}l@{}}- mean tonic\\ - phasic AUC \\ - phasic min, max, sum peaks amplitudes\end{tabular} \\
BVP & \begin{tabular}[c]{@{}l@{}}- phasic min, max, sum peaks amplitudes\\ - mean peak amplitude (diff. between baseline and task)\end{tabular} \\
HR & \begin{tabular}[c]{@{}l@{}}- mean peak amplitude (diff. between baseline and task) \\ - heart-rate variance (diff. between baseline and task)\end{tabular} \\
\hline
\end{tabular}
\caption{Machine learning features grouped by physiological signal and by sensor.}
\vspace{-5mm}
\label{tab:features}
\end{table}
After signals pre-processing, we extracted the features presented in Table~\ref{tab:features}, which we use to train our classifiers. 
We select features based on previous studies using the same signals for machine learning~\cite{MF15,Girardi:ACII2017,FucciGNQL19}.

\textit{\textbf{Classification Settings}}.
In line with previous research on biometrics~\cite{FucciGNQL19,MF15,KoelstraMSLYEPNP12}, we choose eight popular machine learning algorithms---i.e., Naive Bayes (nb), K-Nearest Neighbor (knn), C4.5-like trees (J48), SVM with linear kernel (svm), Multi-layer Perceptron for neural network (mlp), and Random Forest (rf).

We evaluate our classifiers in two different settings.
In the \textit{Hold-out} setting, we split the gold standard into training (90\%) and test (10\%) sets using the stratified sampling strategy implemented in the \textit{R} \texttt{caret} package~\cite{Kuhn09thecaret}. 
We search for the optimal hyper-parameters~\cite{TMH16, TMH18} using leave-one-out cross validation---i.e., the recommended approach for small training sets~\cite{Raschka}, such as ours.
The resulting model is then evaluated on the held-out test set to assess its performance on unseen data.
We repeat this process 10 times to further increase the validity of the results.
The performance is then evaluated by computing the mean for precision, recall, F-measure, and accuracy over the different runs.
This setting is directly comparable to the one implemented by M\"uller and Fritz~\cite{MF15}, which includes data from the same subject in both training and test sets.

We report a second evaluation setting to assess the classifiers performance on data obtained from unseen developers---i.e., \textit{leave-one-subject-out} (LOSO).
This setting was inspired by previous findings reporting different classification performance due to differences in biometrics between individuals~\cite{MF15}.
In this setting, the evaluation on a test set is repeated for 23 times---i.e., the number of subjects in our dataset.  
At each iteration, we use all the observations from 
the \textit{n-1} participants (i.e., 22) for training the model, and we test the performance on the remaining one. 


\textit{\textbf{Classification Performance}}.
In Table~\ref{tab:machineLearning}, for each sensor and their combination, we report the classifier with the highest accuracy, together with its precision, recall, and  F-measure.
Moreover, we report the result of a trivial classifier always predicting the majority class (i.e., \textit{negative} for valence and \textit{high} for arousal)~\footnote{The performance for each run for both settings are reported in the 'BestResults' spreadsheet, in the 'Machine Learning' folder of the replication package.}

In the \textit{hold-out} setting, we observe substantial improvements over the baseline classifier.
The \textit{valence} classifier distinguishes between negative and positive emotions with an accuracy of .$72$  using the full set of sensors.
While performance might appear close to the baseline value in terms of accuracy (baseline accuracy = $.68$), looking at precision and recall, we observe that the classifiers behavior is substantially different.
In fact, we observe an increase of $.34$ in precision (from .34 of the baseline to .68 of the classifier) and of $.10$ (from .50 to .60) in recall, resulting in a $.19$ increase of the F1-measure (from .41 to .60). 
The results show that the model trained using the full set of features achieves comparable performance to the one trained using only features extracted from the Empatica E4 device---i.e., EDA, BVP, and HR-related features.  
We observe a small increase in precision with respect to the full device setting (from $.68$ in the full set to $.70$ with Empatica) and a small decrease in recall (from $.60$ to $.59$). 
These results suggest that valence can be reliably detected using only the Empatica E4 wristband. 
The features associated to the Brainlink EEG helmet negligibly impact the classifiers performance.

For \textit{arousal}, our best classifier distinguishes between \textit{high} and \textit{low} emotion activation with an accuracy of $.65$. 
The model trained using the full set of features substantially outperforms the baseline in terms of precision ($+.31$, from $.31$ to $.62$), recall ($+.11$, from $.50$ to $.61$), and F1-measure ($+.21$, from $.38$ to $.59$).
Similarly to what observed for valence, the performance obtained with the full set of sensors is comparable to the one obtained with the Empatica E4 wristband only, which also achieves a better precision.  
\begin{table*}[htb]
\begin{tabular}{l|p{0.4cm}lllll|p{0.4cm}lllll}
\cline{2-13}
\multicolumn{1}{c|}{} & \multicolumn{6}{c|}{\begin{tabular}[c]{@{}c@{}}Hold-out setting \\ Train: 90\% + LOO cross validation\\ Test: 10\% (10 times)\end{tabular}} & \multicolumn{6}{c|}{\begin{tabular}[c]{@{}c@{}}Leave-one-subject out setting\\ Train: all-1 subject + LOO cross validation\\ Test: 1 held-out subject (23 subjects)\end{tabular}} \\ \hline
\multicolumn{13}{|c|}{\textit{Valence}} \\ \hline
\textbf{Devices} & \textbf{Alg.} & \textbf{Prec} & \textbf{Rec} & \textbf{F1} & \textbf{Accuracy} & \textbf{stdev} & \textbf{Alg.} & \textbf{Prec} & \textbf{Rec} & \textbf{F1} & \textbf{Accuracy} & \textbf{stdev} \\
\hline
Full set & knn & .68 (+.34) & .60 (+.10) & .60 (+.19) & .72 (+.04) & .12 & svm & .48 (+.14) & .62 (+.12) & .53 (+.12) & .69 (+.01) & .25 \\
Empatica & knn & .70 (+.36) & .59 (+.09) & .59 (+.18) & .71 (+.03) & .07& svm & .45 (+.11) & .61 (+.11) & .50 (+.09) & .68 (--) & .27 \\
Brainlink & rf & .54 (+.20) & .54 (+.04) & .52 (+.11) & .66 (-.02) & .07 & mlp & .66 (+.32) & .64 (+.14) & .64 (+.23) & .71 (+.03) & .22\\
\hline
\multicolumn{2}{l}{\textit{Baseline}} & \textit{.34} & \textit{.50} & \textit{.41} & \textit{.68} & \multicolumn{4}{c}{--}\\ \hline
\multicolumn{13}{|c|}{\textit{Arousal}} \\ \hline
Full set & rf & .62 (+.31) & .61 (+.11) & .59 (+.21)  & .65 (+.04) & .05 & svm & .46 (+.15) & .59 (+.09) & .50 (+.12) & .61 (+.05) & .25\\
Empatica & knn & .67 (+.36) & .58 (+.08) & .55 (+.17) & .65 (+.04) & .10 & J48 & .40 (+.09) & .59 (+.09) & .49 (+.11) & .62 (--) & .25\\
Brainlink & rf & .66 (+.35) & .59 (+.09) & .58 (+.20) & .63 (+.01) & .12 & nb & .62 (+.31) & .63 (+.13) & .61 (+.23) & .63 (+.01) & .17\\
\hline
\multicolumn{2}{l}{\textit{Baseline}} & \textit{.31} & \textit{.50} & \textit{.38} & \textit{.62} & \multicolumn{7}{c}{--}\\
\hline
\end{tabular}
\caption{Best valence and arousal classifiers performance. Improvement over the baseline reported in parenthesis.}
\vspace{-5mm}
\label{tab:machineLearning}
\end{table*}

The \textit{LOSO} setting results are similar, for both valence and arousal, to the ones reported in the \textit{hold-out} settings.
However, we observe variability for the individual performance on each test set as suggested by the higher standard deviation compared to the hold-out setting.
These results provide evidence that biometrics are good predictors for emotions, although we observed variability between individuals.

\begin{mdframed}[backgroundcolor=lightgray!20]
\textit{Summary RQ3} - Developers' emotions during programming can be recognized using features extracted by the Empatica E4 wristband (i.e., EDA, BVP, and HR). 
\end{mdframed}

\section{Discussion}
\label{sec:discussion}
In this section, we compare our findings with related studies, 
highlight their implications for researchers and practitioners, and report the threats to their validity. 

\subsection{Comparison with Related Studies}
\label{sec:compareMFGraz}

\textit{\textbf{Emotions as a proxy for progress.}}
The analysis of the scores reported by the participants in our study during the programming task shows a prevalence of negative valence and high arousal. 
This result contrasts with the findings of M\"uller and Fritz~\cite{MF15} who observed that the distribution of emotions reported when programming is comparable to the one reported when watching emotion-triggering pictures.
The prevalence of emotion with negative valence and high arousal in our study can be explained by our participants being less experienced.
In previous work, M\"antyl\"a et al.~\cite{MantylaADGO16} presents evidence that novice developers are more inclined to negative valence and high arousal. 
Furthermore, experience is negatively correlated with effort---i.e., more experienced developers need less effort to complete a task~\cite{Kuutila:survey,Mantyla:2014:ICSE:TimePressure}. 

The lower level of experience of our participants can be seen in their actual and perceived progress.
In fact, the majority reported being either stuck or completely stuck (63\% of self-report questionnaires filled-in during the interruptions).
Conversely, they reported either being in flow or neutral in only 11\% and 26\% of cases, respectively.
These results are consistent with the fact that only 4 over 23 participants completed the task. 
In contrast, M\"uller and Fritz~\cite{MF15} reports a more balanced distribution of progress, with the majority of participants feeling in flow (39\%) rather than stuck (37\%) or neutral (24\%). 

The results of the linear mixed model in our study are comparable to those reported in M\"uller and Fritz~\cite{MF15}. 
We confirm that valence is positively correlated with perceived progress and that it is the main variable explaining the model deviance.
Moreover, we did not show a significant relationship between productivity and arousal. 
Both results confirm a previous study by  Graziotin et al.~\cite{GraziotinWA15:Journal}. 
The positive relationship between valence and progress is consistent with the findings of M\"antyl\"a et al.~\cite{MantylaADGO16} who observed positive emotions when resolving issues in the tracking system (i.e., emotions as a proxy for progress). 
The same study shows low variability in arousal supporting the lack of correlation between this emotional dimension and progress observed in our study. 

\textit{\textbf{Causes for negative emotions and coping strategies}}.  
We confirm previous evidence on the causes of negative emotions and how developers deal with them to regain focus and positive emotions. 
Being stuck and working under time pressure emerged as the most frequent causes for negative emotions. 
Fear of failure was already reported as a cause for frustration in software development~\cite{FordP15:CHASE}.
Similarly, the detrimental impact of limited time on self-confidence, well-being, and emotional states was already observed~\cite{Kuutila:survey,FordP15:CHASE}.

Technical difficulties (e.g., unexpected usage of libraries or unexpected output of code) and unfulfilled information needs (e.g., unavailable documentation) also emerge as causes for negative feelings.
This is consistent with previous investigations of emotions~\cite{FordP15:CHASE, MF15} and confusion~\cite{Ebert:SANER:2019} in software development.
Previous findings suggest that early detection of confusion is crucial for preventing burnout and loss of productivity.
Moreover, they demonstrates how confusion arises due to lack of documentation~\cite{Ebert:SANER:2019}, in presence of unexpected code behavior~\cite{Ebert:SANER:2019}, and  bugs~\cite{MantylaADGO16}. 

We also show that facing new challenges is a trigger for positive emotions, in line with previous work showing that the development of new features causes more positive emotions than bug fixing~\cite{MantylaADGO16}.
Similarly, having new ideas and being in flow while programming is shown to be associated with positive emotions~\cite{MF15}. 



\textit{\textbf{A Minimal Set of Biometrics for Emotion Classification.}}
On top of confirming M\"uller and Fritz~\cite{MF15} findings regarding the usage of non-invasive sensors for valence recognition, we also addressed the classification of the arousal dimension. 
As a novel finding, we identified the minimum set of sensors---EDA, BVP, and HR measured using the Empatica E4 wristband---that can be used in an experimental protocol for detecting emotions during software development tasks. 

Using machine learning, we are able to distinguish between positive and negative valence.
Using only the Empatica E4 wristband, the performance are comparable to ones obtained using the full sensors settings (i.e., wristband + EEG helmet).
Our accuracy ($.72$) is comparable to the one ($.71$) reported by M\"ueller and Fritz ~\cite{MF15}.  
However, their results are obtained using features from an EEG helmet in combination with HR, and pupil size captured by an eye-tracker. 

For arousal classification, our best classifier achieves an accuracy of $.65$ using only features from the Empatica E4 wristband. 
The accuracy of our classifiers is comparable to the one ($.58$) reported by Koelstra et al.~\cite{KoelstraMSLYEPNP12} for arousal classification using a 32-electrode EEG helmet. 
Moreover, they show an accuracy of $.61$ for valence by combining EDA, EMG on facial muscles, features derived from respiration, blood pressure, and eye blinking rate.
We outperform their classifiers using a minimal set of features obtained using the Empatica E4 wristband.

Compared to ours, other studies show better performance---e.g., accuracy for arousal of $.97$~\cite{Soleymani:TAFCC:SAM,Chen,Garcia} and $.91$ for valence~\cite{nogueira2013hybrid}. 
However, these studies rely on high-definition EEG helmets~\cite{Soleymani:TAFCC:SAM,Chen,Garcia} and facial electrodes for EMG \cite{nogueira2013hybrid}. 
Such sensors are invasive and cannot be used outside of a laboratory setting---e.g., in the work environment.

\subsection{Implications}
\textbf{\textit{Implication for researchers}}
The results of this study provides evidence that we can recognize developers' emotions, while programming, by means of a minimal set of biometric features using the sensors mounted on a single wearable device (the Empatica E4 wristband).
This opens up the possibility of further studies aimed at improving the ecological validity of our findings.

Our results show between-subject variability of biometrics, already observed in previous studies~\cite{MF15}. 
The higher standard deviation for accuracy in the LOSO setting can be problematic when classifying the emotions of a new unseen developer.
Further studies with a wider pool of participants are required to assess the robustness of our valence and arousal classifiers. 
Such studies can investigate to what extent we can build more robust classifiers by performing preliminary subject-based calibration---e.g., by tuning the models based on individuals' biometrics collected while exposing participants to emotional stimuli in a controlled setting. 
Future studies can identify the amount of biometric data required to fine-tune the models for a reliable classification of emotions of new subjects.

The correlation between valence and progress can be interpreted as a proxy for self-perceived productivity.
However, further investigation is required to \textit{(i) } provide an explanation for the results of our correlation study, and \textit{(ii)} understand whether a causal relation exists between emotions and productivity (or vice-versa)---e.g., using emotion-triggering techniques in a controlled setting, in line with previous research~\cite{Khan2011}.

\textbf{\textit{Implication for practitioners and tool builders}}.
We show that emotions can be detected using non-invasive wearable device, such as a wristband.
This finding paves the way for tools and practices to prevent developers' distress and burnout.

Early recognition of negative emotions, integrated with the development environment, can be leveraged to suggest corrective actions.
Developers can regain focus and restore positive moods in accordance with the strategies we observed in this study to cope with negative emotions.

Our results demonstrate a positive correlation between valence and progress, suggesting that emotions might act as a proxy for productivity.
For example, positive emotions can indicate that a developer is \textit{in flow} and should not be disturbed. 
Hence, sensor-based emotion classifiers can improve state-of-the-art approaches for the automatic assessment of interruptibility~\cite{ZugerMMF18}.
Similarly, the identification of negative emotions can indicate a stuck developer requiring support fulfilling her information needs.
Accordingly, an emotion-aware component integrated in the development environment can recommend relevant colleagues to consult on the code base~\cite{Kavaler} or trigger utilities for on-demand documentation generation~\cite{OD3}. 

Biometrics can enhance retrospective meetings by including emotional information collected day-to-day rather than at the end of an iteration or sprint.
The team can better identify what are the activities and events that relate to positive and negative emotions~\cite{Girardi:2019}.

\subsection {Threats to Validity}
In this section, we report the threats to the validity in increasing order of priority for the \textit{in vitro} nature of this study, following the recommendations of Wohlin et al.~\cite{WRH12}.
\label{sec:threats}

\textbf{\textit{External validity}}
Threats to external validity relate to the generalizability of the results. 
We chose the same task used in a former study~\cite{MF15}, which simulated a new problem in a real scenario.
Regarding participants, we covered different levels of academic experience (by including Bachelors, Masters, and PhD students) but with less professional experience compared to~\cite{MF15}.

\textbf{\textit{Conclusion validity}}
The validity of our conclusions relies on the robustness of the generalized linear model and machine learning models. 
We mitigated such threat by \textit{(i)} running several algorithms addressing the same classification task, \textit{(ii)} applying hyper-parameters tuning to optimally solve the task, and i\textit{(iii)} reporting results from two different evaluation settings---i.e., Hold-out and LOSO.

\textbf{\textit{Construct validity}}
Our study suffers from threats to construct validity---i.e., the reliability of our measures in capturing emotions and progress.
When assessing the impact of arousal on progress, we did not observe a significant correlation. 
Although we cannot exclude that such result is due to the unreliability of the self-reported rating, we performed data quality assurance and did not consider participants who misinterpreted the concept of arousal---e.g., who reported always the same score also during emotion elicitation. 

\textbf{\textit{Internal validity}}
Threats to internal validity concern confounding factors that can influence the results. 
We collected data in a laboratory setting.
Factors existing in our settings, such as the presence of the experimenter and the absence of real consequences when failing or succeeding in the task, can influence the triggered emotions---e.g., negative emotions due to the feeling of being observed or judged.
In addition, interrupting developers during the task can have interfered with their work and elicited negative emotions.
We mitigate this threat by interrupting the developers when we notice a task switch (e.g., opening a new browser window) in proximity of the five-minutes interval.
\section{Conclusion}
\label{sec:conclusion}
We investigated the range and triggers of emotions experienced by software developers during a programming task.
We confirm the link between emotions and self-reported progress observed in previous studies. 

Using EDA and heart-related metrics collected using a wristband, we trained a machine learning classifier that can accurately recognize valence.
A second classifier, trained to recognized arousal, has shown less successful but encouraging results. 




Our results can stimulate future \textit{in-vivo} research (i.e., in software development companies) and the collection of biometrics to explore emotions during the entire working day, when developers are involved in different activities, not just programming.
Furthermore, observing developers at the workplace also opens opportunities to build more sophisticated classifiers, which are able to identify, for example, the bad days (i.e., when mostly negative emotions are identified) or negative working conditions of developers (i.e., when negative emotions are observed over a long period of time).

\bibliographystyle{ACM-Reference-Format}
\bibliography{sample-base}

\end{document}